\documentclass[12pt]{article}
\usepackage{oldgerm}
\usepackage{yfonts}
\usepackage{euler}
\usepackage{graphics}
\usepackage{bm}
\usepackage{graphicx}
\usepackage{amsmath}
\usepackage{amssymb}
\usepackage{amstext}
\usepackage{amscd}
\usepackage{amsfonts}
\usepackage{appendix}
\usepackage{color}
\usepackage{wasysym}
\usepackage{latexsym}
\usepackage{epstopdf}

\DeclareMathAlphabet{\EuFrak}{U}{euf}{m}{n}
\DeclareMathAlphabet{\EuScript}{U}{eus}{m}{n}

\newcommand{\nd}{\noindent}

\newcommand{\be}{\begin{equation}}
\newcommand{\ee}{\end{equation}}
\newcommand{\ben}{\begin{eqnarray}}
\newcommand{\een}{\end{eqnarray}}

\title{{\bf Rescuing the MaxEnt treatment for
$q$-generalized entropies}}

\author{{A. Plastino$^{1,2,4}$, M.C.Rocca$^{1,2,3}$}, \\
\small{$^1$ Departamento de F\'{\i}sica,
Universidad Nacional de La Plata,}\\
\small{$^2$ Departamento de Matem\'{a}tica,
Universidad Nacional de La Plata,}\\
\small{$^3$ Consejo Nacional de Investigaciones Cient\'{\i}ficas
y Tecnol\'{o}gicas}\\
\small{(IFLP-CCT-CONICET)-C. C. 727, 1900 La Plata -
Argentina}\\\small{$^4$  SThAR - EPFL, Lausanne, Switzerland}}

\date{\today}

\begin{document}

\maketitle

\begin{abstract}

It has been recently argued, via very clever arguments, that the
MaxEnt variational problem would not adequately work for Renyi's and
Tsallis' entropies. We constructively show here that this is not so
if one formulates the associated variational problem in a more
orthodox functional fashion.\\
%\nd PACS: 05.30.-d, 05.20-y, 05.70.-a\\
\nd {\bf Keywords:  Tsallis entropy, Renyi entropy, MaxEnt}

\end{abstract}

\newpage

\renewcommand{\theequation}{\arabic{section}.\arabic{equation}}

\setcounter{equation}{0}

\section{Introduction}

 Renyi's and Tsallis' information measures ($S_R$ and $S_T$), respectively) are generalizations of
 the Shannon one, quantifying a system's
diversity, uncertainty, or randomness. The degree of generalization is parameterized by a real parameter.
Both are very important
quantities  in variegated areas of scientific research. One can
mention, for instance,  ecology, quantum information,  the
Heisenberg XY spin chain model, theoretical computer science,
diffusion processes, several biological processes, high energy physics,
etc. As a small sample, see for example,
\cite{1,2,3,4,5,6,7,8,9,10}, \cite{tsallis,web,pre,epjb1,epjb2,epjb3,epjb4,epjb5,epjb6,epjb7,epjb8,epjb9}.

\vskip 3mm \nd Information Theory (IT)  yields a quite powerful
inference methodology,  abbreviated as MaxEnt \cite{jaynes}, that is
able to describe  general properties of arbitrary systems, in most
areas the  of scientific endeavor,  on the basis of scarce
information. MaxEnt  yields the least-biased description that can be
gotten according to specific data-amounts, in any possible
circumstances  \cite{jaynes}. In
 statistical mechanics (SM), Jaynes  pioneered the
use of MaxEnt so as to both i) reformulate SM and ii)
generalize its foundations  \cite{jaynes}. Here we focus on the MaxEnt technique as applied
to Renyi-Tsallis information measures.

\vskip 3mm   \nd  A very interesting result was recently reported in \cite{bagci}. Let $a$ and $\beta$ stand for
the two Lagrange multipliers appearing in a canonical ensemble MaxEnt treatment. Then, one confidently expects
 MaxEnt to yield, for the canonical probability distribution (PD) $P$ ($Z$ being the partition function,
$f$ an appropriate function, and $\epsilon$ the level energy),

\be \label{uno} P_i= f(a + \beta \epsilon_i)/Z,\ee

\be \label{dos} Z= \sum_i\,f(a + \beta \epsilon_i).\ee It is claimed
in \cite{bagci}, by recourse to an ingenious procedure, that for
this to be possible $f$ must be endowed with certain properties.
These properties, it is shown in \cite{bagci}, are NOT possessed by
the PDs that maximize either Tsallis' or Renyi's entropies.  This
constitutes an intriguing result that deserves further
consideration. \vskip 3mm

\nd We show here by  construction, appealing to conventional
techniques of Functional Analysis, not used in \cite{bagci}, that
the {\it functional forms}  (\ref{uno}),  (\ref{dos}) are indeed recovered \`a la
Tsallis (or Renyi), thus surmounting the formidable obstacle posed
by \cite{bagci} to the use of MaxEnt for these two entropies. We are
not claiming that the arguments of \cite{bagci} are wrong, but that
they are Calculus-based, while the proper MaxEnt scenario requires
functional Analysis. The central notion required here is that of
increment $ h$ of a functional. The general theory of Variational
Calculus is concerned with Banach Space (BS) \cite{tp2}, examples of
which are Hilbert's space and classical phase space. The MaxEnt
technique in Banach space demands a first variation that has to
vanish and a second one that ascertains the characteristics of the
ensuing extremum.  The main need is then  to evaluate the increment
$h$ of a functional $F$ at the point $x$ of the operative BS.   {\it
One must remember that functional calculus is not identical to
ordinary calculus (involving ordinary functions), particularly
when one is looking for extremes \cite{tp1,tp2}.}%%%%%%%%%%%%%%%%%%%%%%%%\vskip 3mm

\setcounter{equation}{0}

\section{Tsallis MaxEnt Treatment}
 The concomitant Tsallis' procedure has been developed in \cite{tp1}.   The
Tsallis information   functional is \cite{tsallis}
\begin{equation}
\label{eq3.1}
F_S(P)=-\sum\limits_{i=1}^n
P_i^q\ln_q(P_i)+\lambda_1\left(
\sum\limits_{i=1}^{n}P_iU_i-<U>\right)+\lambda_2
\left(\sum\limits_{i=1}^{n}P_i-1\right)
\end{equation}
For the increment one has
\[F_S(P+h)-F_S(P)=-\sum\limits_{i=1}^n(P_i+h_i)^q\ln_q(P_i+h_i)
+\lambda_1\left[
\sum\limits_{i=1}^n(P_i+h_i)U_i-<U>\right]+\]
\[\lambda_2\left[\sum\limits_{i=1}^n(P_i+h_i)-1\right]+
\sum\limits_{i=1}^nP_i^q\ln_q(P_i)-\lambda_1\left(
\sum\limits_{i=1}^nP_iU_i-<U>\right)-\]
\begin{equation}
\label{eq3.2}
\lambda_2\left(\sum\limits_{i=1}^nP_i-1\right)
\end{equation}
Eq.  (\ref{eq3.2}) can be rewritten  as
\[F_S(P+h)-F_S(P)=
\sum\limits_{i=1}^n\left[\left(\frac {q} {1-q}\right)P_i^{q-1}
+\lambda_1 U_i +\lambda_2\right]h_i\]
\begin{equation}
\label{eq3.3} \sum\limits_{i=1}^nqP_i^{q-2}\frac {h_i^2}
{2}+O(h^3),
\end{equation}
Eq. (\ref{eq3.3}) leads to
\begin{equation}
\label{eq3.4} \left(\frac {q} {1-q}\right)P_i^{q-1} +\lambda_1 U_i
+\lambda_2=0,
\end{equation}
\begin{equation}
\label{eq3.5} -\sum\limits_{i=1}^nqP_i^{q-2}h_i^2\leq C||h||^2
\end{equation}
Eq. (\ref{eq3.4}) is the  Euler-Lagrange one while (\ref{eq3.5})
gives bounds originating from the second variation.  Thus,
(\ref{eq3.4}) entails
\begin{equation}
\label{eq3.6} \lambda_1=\beta q  Z^{1-q}
\end{equation}
\begin{equation}
\label{eq3.7}
\lambda_2=\frac {q} {q-1}Z^{1-q}
\end{equation}

\nd The following two basic relations for our task follow now, namely,
\begin{equation}
\label{eq3.8}
P_i=\frac {[1+\beta(1-q)U_i]^{\frac {1} {q-1}}} {Z}
\end{equation}
\begin{equation}
\label{eq3.9}
Z=\sum\limits_{i=1}^n [1+\beta(1-q)U_i]^{\frac {1}
{q-1}}.
\end{equation}
These two relations should not exist according to  \cite{bagci}.
 That we can construct them follows from Functional Analysis, that was not appealed to in that reference.

\setcounter{equation}{0}

\section{Renyi's MaxEnt Treatment}

Renyi's $S_R$ is defined as \cite{9}:
\begin{equation}
\label{eq2.1} S_R=\frac {1} {1-\alpha}\ln\left( \sum\limits_{i=1}^n
P_i^{\alpha}\right),
\end{equation}
and the accompanying (canonical ensemble) MaxEnt probability
distribution $P$ arises from the maximization of the functional
$F_{S_R}(P)$ [where $U$ denotes the energy and $<U>$ its mean
value]
\begin{equation}
\label{eq2.2} F_{S_R}(P)=\frac {1} {1-\alpha}\ln\left(
\sum\limits_{i=1}^n P_i^{\alpha}\right)+ \lambda_1\left(\sum\limits_{i=1}^n
P_iU_i-<U>\right)+ \lambda_2\left(\sum\limits_{i=1}^n P_i-1\right),
\end{equation}
 Following
standard {\it Functional Analysis} procedures (which are  not employed in Ref- \cite{bagci},
  we consider the functional-$h$ increment
\cite{tp1,tp2}
\[F_{S_R}(P+h)=\frac {1} {1-\alpha}\ln\left[
\sum\limits_{i=1}^n
(P_i+h_i)^{\alpha}\right]+
\lambda_1\left[\sum\limits_{i=1}^n
(P_i+h_i)U_i-<U>\right]+\]
\begin{equation}
\label{eq2.3} \lambda_2\left[\sum\limits_{i=1}^n (P_i+h_i)-1\right],
\end{equation}
so that
\[F_{S_R}(P+h)-F_{S_R}(P)=\frac {1} {1-\alpha}\ln\left[
\sum\limits_{i=1}^n
(P_i+h_i)^{\alpha}\right]-
\frac {1} {1-\alpha}\ln\left(
\sum\limits_{i=1}^n
P_i^{\alpha}\right)+\]
\begin{equation}
\label{eq2.4} \lambda_1\sum\limits_{i=1}^n h_iU_i+\lambda_2
\sum\limits_{i=1}^n h_i.
\end{equation}
We now tackle $h^2$ contributions  so as to assess second
variations of $F_{S_R}$  \cite{tp1,tp2}
\[F_{S_R}(P+h)-F_{S_R}(P)=\frac {1} {1-\alpha}\ln\left\{
\sum\limits_{i=1}^n
\left[P_i^{\alpha}+
\alpha h_iP_i^{\alpha-1}+
\frac {\alpha(\alpha-1)} {2}
h_i^2P_i^{\alpha-2}
\right]\right\}-\]
\begin{equation}
\label{eq2.5} \frac {1} {1-\alpha}\ln\left( \sum\limits_{i=1}^n
P_i^{\alpha}\right)+ \lambda_1\sum\limits_{i=1}^n h_iU_i+\lambda_2
\sum\limits_{i=1}^n h_i.
\end{equation}
or,  equivalently,
\[F_{S_R}(P+h)-F_{S_R}(P)=\frac {1} {1-\alpha}\ln\left\{ 1+
\frac {\sum\limits_{i=1}^n
\left[\alpha h_iP_i^{\alpha-1}+
\frac {\alpha(\alpha-1)} {2}
h_i^2P_i^{\alpha-2}
\right]}
{\sum\limits_{i=1}^n
P_i^{\alpha}}
\right\}+\]
\begin{equation}
\label{eq2.6} \lambda_1\sum\limits_{i=1}^n h_iU_i+\lambda_2
\sum\limits_{i=1}^n h_i,
\end{equation}
so that one finally arrives at

\[F_{S_R}(P+h)-F_{S_R}(P)=\frac {1}
{1-\alpha} \frac {\sum\limits_{i=1}^n \left[\alpha h_iP_i^{\alpha-1}+ \frac
{\alpha(\alpha-1)} {2} h_i^2P_i^{\alpha-2} \right]} {\sum\limits_{i=1}^n
P_i^{\alpha} }+\]
\begin{equation}
\label{eq2.7}
-\frac {1} {2(1-\alpha)}\left(
\frac {\sum\limits_{i=1}^n \alpha h_i P_i^{\alpha-1}}
{\sum\limits_{i=1} P_i^{\alpha}}\right)^2+
\lambda_1\sum\limits_{i=1}^n h_iU_i+\lambda_2
\sum\limits_{i=1}^n h_i.
\end{equation}
Summing up, we have for the first variation

\begin{equation}
\label{eq2.8} \frac {\alpha} {1-\alpha} \frac {P_i^{\alpha-1}}
{\sum\limits_{i=1}^n P_i^{\alpha}}+\lambda_1 U_i+\lambda_2=0.
\end{equation}

Multiplying (\ref{eq2.8})  by  $P_i$ we find
\begin{equation}
\label{eq2.10} \frac {\alpha} {1-\alpha} \frac {P_i^{\alpha}}
{\sum\limits_{i=1}^n P_i^{\alpha}}+\lambda_1 P_iU_i+\lambda_2P_i=0.
\end{equation}
Integrating now we are led to
\begin{equation}
\label{eq2.11} \frac {\alpha} {1-\alpha} +\lambda_1<U>+\lambda_2=0.
\end{equation}
This is an important result, showing that  $\lambda_1$ and
$\lambda_2$ are not  independent Lagrange multipliers. We are authorized  to write
\begin{equation}
\label{eq2.12} \lambda_2= \frac {\alpha} {\alpha-1} -\lambda_1<U>,
\end{equation}
and replacing this value of  $\lambda_2$ in (\ref{eq2.8}) we get

\begin{equation}
\label{eq2.13} \frac {\alpha} {1-\alpha} \frac {P_i^{\alpha-1}}
{\sum\limits_{i=1}^n P_i^{\alpha}}+\lambda_1(U_i-<U>)+ \frac {\alpha}
{\alpha-1}=0,
\end{equation}
whose solution is given by

\begin{equation}
\label{eq2.14}
\lambda_1=-\beta\alpha
\end{equation}

\begin{equation}
\label{eq2.15}
P_i=\frac {[1+\beta(1-\alpha)(U_i-<U>)]^{\frac {1} {\alpha-1}}} {Z}
\end{equation}

\begin{equation}
\label{eq2.16}
Z=\sum\limits_{i=1}^n [1+\beta(1-\alpha)(U_i-<U>)]^{\frac {1}
{\alpha-1}}.
\end{equation}

\nd Ref. \cite{bagci} showed, via an alternative, calculus-based  procedure, that Eqs. (\ref{eq2.15}) and (\ref{eq2.16})
are impossible to obtain in this form for Renyi's entropy. We are able here to contradict such assertion because we employed Functional Analysis and not just ordinary calculus. Finally, note that Eqs. (8) and (11) of
\cite{bagci} {\it define} both the exponential and the logarithm functions. Thus, their co-implication is obvious.

\setcounter{equation}{0}

\section{Conclusions}

The MaxEnt variational treatment in the canonical ensemble is believed to yield, for the probability
distribution   (PD) $P$ that maximizes the entropy the forms

\be \label{tres} P_i= f(a + \beta \epsilon_i)/Z,\ee

\be \label{cuatro} Z= \sum_i\,f(a + \beta \epsilon_i),\ee
where $a$ and $\beta$ are Lagrange multipliers, $Z$ is the partition function, $f$ is an appropriate function,
 and $\epsilon$ is a level-energy.

It is claimed in \cite{bagci}, by recourse to an admirable,
Calculus-based procedure, that for this to be possible $f$ must be
endowed with special properties. These properties, one finds in
\cite{bagci}, are NOT possessed by the PDs that maximize either
Tsallis' or Renyi's entropies. \vskip 3mm

\nd We have shown here (by  construction), appealing to conventional techniques of Functional Analysis, not used in
\cite{bagci}, that the {\it functional forms} (\ref{uno}),  (\ref{dos}) are indeed encountered in dealing with Tsallis' (or Renyi's) entropies. This removes the rather formidable obstacle posed by \cite{bagci} to the use of MaxEnt for these two entropies. Note that we are not claiming that the arguments of \cite{bagci} are wrong, but that they are Calculus-based, while the proper MaxEnt scenario requires functional Analysis.

\end{document}